\documentclass[aps,pra,10pt,superscriptaddress,eqsecnum,showpacs,amsmath,twocolumn,floatfix]{revtex4-1}
\usepackage{amsmath}
\usepackage{amsfonts}
\usepackage{amssymb}
\usepackage{natbib}
\usepackage{graphicx}

\newcommand{\diff}[2]{\frac{d #1}{d #2}}

\newcommand{\intall}{\int_{-\infty}^{\infty}}
\newcommand{\ket}[1]{|#1\rangle}

\newcommand{\bra}[1]{\langle#1|}

\newcommand{\avg}[1]{\langle#1\rangle}
\newcommand{\Avg}[1]{\left\langle#1\right\rangle}

\newcommand{\abs}[1]{\left|#1\right|}
\newcommand{\bk}[1]{\left(#1\right)}
\newcommand{\Bk}[1]{\left[#1\right]}
\newcommand{\BK}[1]{\left\{#1\right\}}

\newcommand{\trace}{\operatorname{tr}}

\newcommand{\erfc}{\operatorname{erfc}}

\begin{document}
%\twocolumn[
\title{Fundamental quantum limits to waveform detection}

\author{Mankei Tsang}

\email{eletmk@nus.edu.sg}
\affiliation{Department of Electrical and Computer Engineering,
  National University of Singapore, 4 Engineering Drive 3, Singapore
  117583}

\affiliation{Department of Physics, National University of Singapore,
  2 Science Drive 3, Singapore 117551}

\author{Ranjith Nair}

\affiliation{Department of Electrical and Computer Engineering,
  National University of Singapore, 4 Engineering Drive 3, Singapore
  117583}

%\affiliation{
%Center for Quantum Information and Control,
%University of New Mexico, MSC07--4220, Albuquerque, New Mexico
%87131-0001, USA}

%\author{Carlton M.\ Caves}

%\affiliation{
%Center for Quantum Information and Control,
%Department of Physics and Astronomy,
%University of New Mexico, Albuquerque, New Mexico
%87131, USA}

%\author{Seth Lloyd}

%\affiliation{Research Laboratory of Electronics,
%Massachusetts Institute of Technology, Cambridge, Massachusetts
%02139, USA}

%\affiliation{Department of Mechanical Engineering,
%Massachusetts Institute of Technology, Cambridge, Massachusetts
%02139, USA}

%\author{Jeffrey H.\ Shapiro}

%\affiliation{Research Laboratory of Electronics, Massachusetts
%  Institute of Technology, Cambridge, Massachusetts 02139, USA}

\date{\today}

\begin{abstract}
  Ever since the inception of gravitational-wave detectors, limits
  imposed by quantum mechanics to the detection of time-varying
  signals have been a subject of intense research and debate. Drawing
  insights from quantum information theory, quantum detection theory,
  and quantum measurement theory, here we prove lower error bounds for
  waveform detection via a quantum system, settling the
  long-standing problem. In the case of optomechanical force
  detection, we derive analytic expressions for the bounds in some
  cases of interest and discuss how the limits can be approached using
  quantum control techniques.
\end{abstract}

\maketitle

\section{Introduction}
The study of quantum measurement has come a long way since the
proposal of wavefunction collapse by Heisenberg and von Neumann, the
philosophical debates by Bohr and Einstein, and the cat experiment
hypothesized by Schr\"odinger. With more and more experimental
demonstrations of bizarre quantum effects being realized in
laboratories, many researchers have shifted their focus to the
practical implications of quantum mechanics for precision
measurements, such as gravitational-wave detection, optical
interferometry, atomic clocks, and magnetometry
\cite{schnabel,chu,budker,glm_science}. Braginsky, Thorne, Caves, and
others pioneered the application of quantum measurement theory to
gravitational-wave detectors \cite{braginsky,bvt,caves}, while
Holevo, Yuen, Helstrom, and others have developed a beautiful theory
of quantum detection and estimation \cite{helstrom,holevo} based on
the more abstract notions of quantum states, effects, and operations
\cite{kraus}. Although Holevo \textit{et al.}'s approach was able to
produce rigorous proofs of quantum limits to various information
processing tasks, so far it has been applied mainly to simple quantum
systems with trivial dynamics measured destructively to extract static
parameters. Applying such an approach to gravitational-wave detection,
or optomechanical force detection in general \cite{aspelmeyer}, proved
to be far trickier; the signal of interest there is time-varying
(commonly called a waveform in engineering literature
\cite{vantrees}), the detector is a dynamical system, and the
measurements are nondestructive and continuous
\cite{braginsky,bvt,caves}.  Quantum limits to such detectors had
been a subject of debate \cite{yuen1983,ozawa1988,caves1985}, with no
definitive proof that any limit exists. In more recent years, the
rapid progress in experimental quantum technology suggests that
quantum effects are becoming relevant to metrological applications and
has given the study of quantum limits a renewed impetus
\cite{schnabel,chu,budker,aspelmeyer}.

Generalizing the quantum Cram\'er-Rao bound first proposed by Helstrom
\cite{helstrom}, Tsang, Wiseman, and Caves recently derived a quantum
limit to waveform estimation \cite{twc}, which represents the first
step towards a rigorous treatment of quantum limits to a waveform
sensor. That work assumes that one is interested in estimating an
existing waveform accurately, so that the mean-square error is an
appropriate error measure. The first goal of gravitational-wave
detectors is not estimation, however, but to detect the existence of
gravitational waves, in which case the miss and false-alarm
probabilities are the more relevant error measures \cite{vantrees} and
the existence of quantum limits remains an open problem. Here we
settle this long-standing question by proving lower error bounds for
the quantum waveform detection problem. To illustrate our results, we
apply them to optomechanical force detection, demonstrating a
fundamental trade-off between force detection performance and
precision in detector position, and discuss how the limits can be
approached in some cases of interest using a quantum-noise
cancellation (QNC) technique
\cite{qnc,qmfs,caniard,julsgaard,hammerer,wasilewski} and an
appropriate optical receiver, such as the ones proposed by Kennedy and
Dolinar \cite{helstrom,cook}.  Merging the continuous quantum
measurement theory pioneered by Braginsky \textit{et al.}\ and the
quantum detection theory pioneered by Holevo \textit{et al.}, these
results are envisaged to play an influential role in quantum
metrological techniques of the future.

\section{Quantum detection of a classical waveform}
Let $P[y|\mathcal H_0]$ be the probability functional of an
observation process $y(t)$ under the null hypothesis $\mathcal H_0$,
and 
\begin{align}
P[y|\mathcal H_1] = \int Dx P[x] P[y|x,\mathcal H_1]
\end{align}
be the probability functional under the alternative hypothesis
$\mathcal H_1$. $x(t)$ is a classical waveform, $P[x]$ is its prior
probability functional, and $P[y|x,\mathcal H_1]$ is the likelihood
functional under $\mathcal H_1$. To perform hypothesis testing given a
record of $y(t)$, one separates the observation space into two
decision regions $\Upsilon_0$ and $\Upsilon_1$, such that $\mathcal
H_0$ is chosen if $y$ falls in $\Upsilon_0$ and $\mathcal H_1$ is
chosen if $y$ falls in $\Upsilon_1$.  The miss probability is defined
as 
\begin{align}
P_{01}\equiv \int_{\Upsilon_0} Dy P[y|\mathcal H_1]
\end{align} and the
false-alarm probability is 
\begin{align}
P_{10}\equiv \int_{\Upsilon_1} Dy P[y|\mathcal H_0].
\end{align}
Two popular decision strategies are the Bayes criterion, which
minimizes the average error probability 
\begin{align}
P_e\equiv P_{10}P_0 + P_{01}P_1
\end{align}
given the prior hypothesis probabilities $P_0$ and $P_1 =
1-P_0$, and the Neyman-Pearson criterion, which minimizes $P_{01}$ for
an allowable $P_{10}$, or vice versa \cite{vantrees}.

\begin{figure}[htbp]
\centerline{\includegraphics[width=0.45\textwidth]{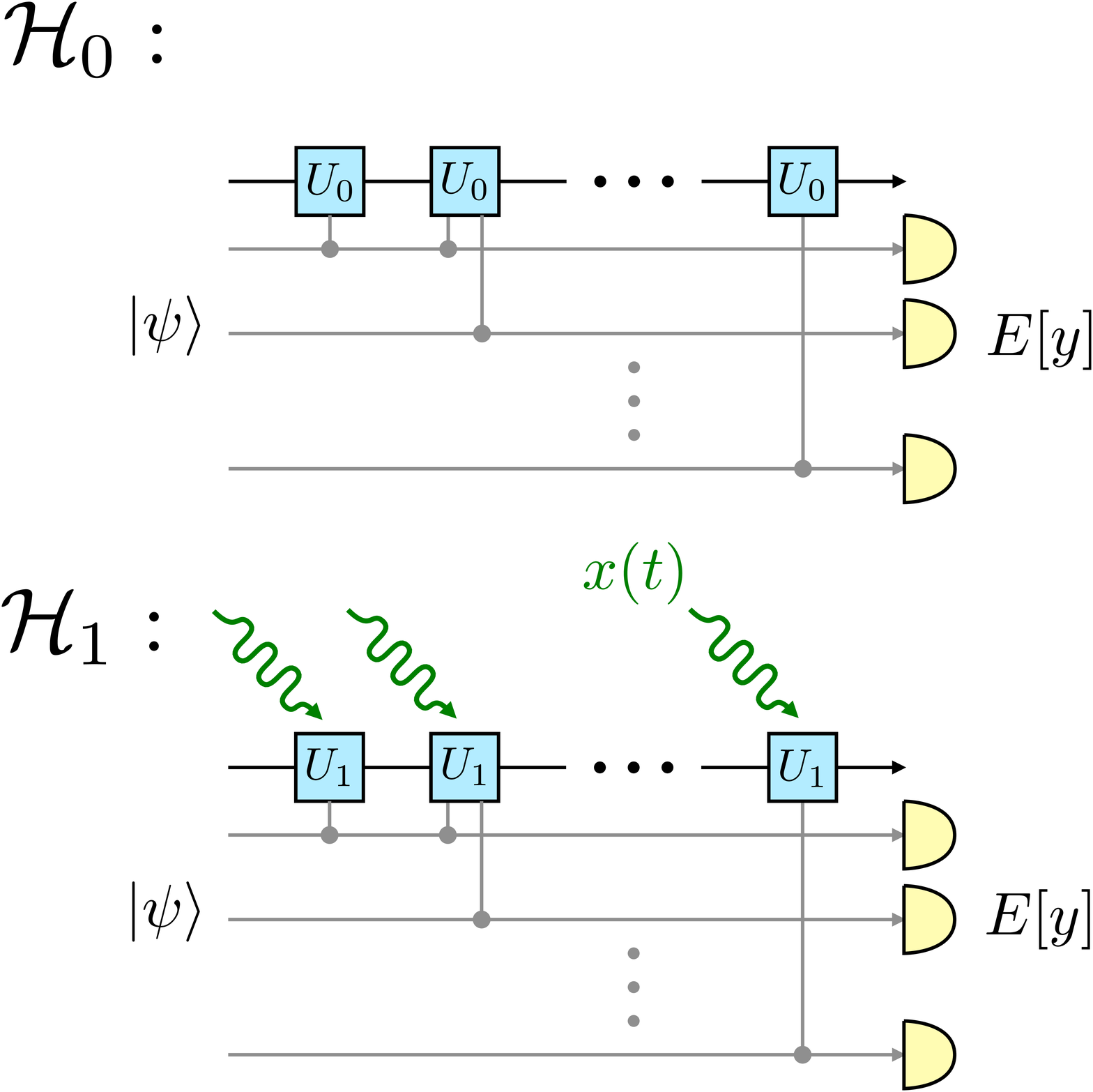}}
\caption{Quantum-circuit diagrams for the waveform detection problem.
  The quantum system is modeled as a pure state $\ket{\psi}$ with
  unitary evolution ($U_0$ or $U_1$) under each hypothesis ($\mathcal
  H_0$ or $\mathcal H_1$) in a large enough Hilbert space for a given
  classical waveform $x(t)$, which perturbs the evolution under
  $\mathcal H_1$. If $x(t)$ is stochastic, the final quantum state
  under $\mathcal H_1$ is mixed. Measurements are modeled as a
  positive-operator-valued measure (POVM) $E[y]$ at the final time
  through the principle of deferred measurement.}
\label{naimark}
\end{figure}

To introduce quantum mechanics to the problem, assume that $x(t)$
perturbs the dynamics of a quantum system under $\mathcal H_1$ and
$y(t)$ results from measurements of the system.  Without any loss of
generality, we model $P[y|\mathcal H_0]$ and $P[y|x,\mathcal H_1]$ by
considering a large enough Hilbert space, such that the initial
quantum state $\ket{\psi}$ at time $t_i$ is pure, the evolution in the
Schr\"odinger picture is unitary, and measurements are modeled by a
positive-operator-valued measure (POVM) $E[y]$ at the the final time
$t_f$ via the principle of deferred measurement
\cite{kraus,twc,nielsen}:
\begin{align}
P[y|\mathcal H_0] &= \trace
\BK{E[y]U_0(t_f,t_i)\ket{\psi}\bra{\psi}U_0^\dagger(t_f,t_i)},
\\
P[y|x,\mathcal H_1] &= \trace
\BK{E[y]U_1(t_f,t_i)\ket{\psi}\bra{\psi}U_1^\dagger(t_f,t_i)},
\end{align}
where
only the unitaries $U_0$ and $U_1$ are assumed to differ and $U_1$
depends on $x$. Assume further that
\begin{align}
U_0(t_f,t_i) &= \mathcal T \exp\Bk{-\frac{i}{\hbar}
\int_{t_i}^{t_f} dt H_0(t)},
\\
U_1(t_f,t_i) &= \mathcal T \exp\Bk{-\frac{i}{\hbar}
\int_{t_i}^{t_f} dt H_1(x(t),t)},
\\
H_1(x(t),t) &= H_0(t) + \Delta H(x(t),t),
\end{align}
where $\mathcal T$ denotes time-ordering and $\Delta H(x(t),t)$ is the
Hamiltonian term responsible for the coupling of the waveform $x(t)$
to the quantum detector.
Figure~\ref{naimark} shows the quantum-circuit
diagrams \cite{caves_shaji} that depict the problem.

This setup can now be cast as a problem of quantum state
discrimination between a pure state 
\begin{align}
\rho_0\equiv
U_0\ket{\psi}\bra{\psi}U_0^\dagger
\end{align}
and a mixed state 
\begin{align}
\rho_1\equiv
\int DxP[x]U_1\ket{\psi}\bra{\psi}U_1^\dagger.
\end{align}
Let $\ket{\Psi_j}$ be a purification of $\rho_j$ in a larger Hilbert
space $\mathbb H_A \otimes \mathbb H_B$, such that $\rho_j = \trace_B
\ket{\Psi_j}\bra{\Psi_j}$ and $\trace\{E[y]\rho_j\} =
\trace[(E[y]\otimes 1_B)\ket{\Psi_j}\bra{\Psi_j}]$, where $1_B$
denotes the identity operator with respect to $\mathbb H_B$.  The
average error probability is thus lower-bounded by \cite{helstrom}:
\begin{align}
P_e &\ge \frac{1}{2}\bk{1-\sqrt{1-4P_0P_1|\avg{\Psi_0|\Psi_1}|^2}},
\end{align}
which is valid for any purification. Hence
\begin{align}
P_e &\ge \frac{1}{2}\bk{1-\sqrt{1-4P_0P_1 \max_{\ket{\Psi_0},\ket{\Psi_1}}
\abs{\avg{\Psi_0|\Psi_1}}^2}}
\\
&= \frac{1}{2}\bk{1-\sqrt{1-4P_0P_1 F}},
\label{Pe}
\end{align}
where $F$ is the quantum fidelity by Uhlmann's theorem \cite{nielsen}:
\begin{align}
F(\rho_0,\rho_1) &\equiv \bk{\trace \sqrt{\sqrt{\rho_1}\rho_0\sqrt{\rho_1}}}^2.
\end{align}
As $\rho_0$ is pure, the fidelity is given by
\begin{align}
F &= \bra{\psi}U_0^\dagger \rho_1 U_0\ket{\psi} = \mathbb E(F_x),
\\
F_x &\equiv \abs{\Avg{U_0^\dagger(t_f,t_i) U_1(t_f,t_i)}}^2,
\label{Fx}
\end{align}
where we have defined classical and quantum averages by
\begin{align}
\mathbb E (\cdot) &\equiv \int Dx P[x](\cdot),
\\
\avg{\cdot} &\equiv \bra{\psi}\cdot\ket{\psi}.
\end{align}
By similar arguments, a quantum bound on the miss probability $P_{01}$
for a given allowable false-alarm probability $P_{10}$ can be derived
from the bound for the pure-state case \cite{helstrom}:
\begin{align}
P_{01} 
&\ge \Big\{\begin{array}{ll}
1-[\sqrt{P_{10}F}+\sqrt{(1-P_{10})(1-F)}]^2, &
P_{10} \le F;
\\
0, & P_{10} \ge F.
\end{array}
\label{P01}
\end{align}
Note that the latter bound is equally
valid if we interchange $P_{01}$ and $P_{10}$; for example, fixing
$P_{01} = 0$ means $P_{10} \ge F$. Equations~(\ref{Pe}) and
(\ref{P01}) are valid for any POVM and achievable if $x(t)$ is known
\textit{a priori}, such that both $\rho_0$ and $\rho_1$ are pure
\cite{helstrom}. 

In terms of related prior work at this point, Ou \cite{ou} and Paris
\cite{paris} studied quantum limits to interferometry in the context
of detection, while Childs \textit{et al.}\ \cite{childs}, Ac\'in
\textit{et al.}\ \cite{acin2001,ajv}, and D'Ariano \textit{et al.}\
\cite{dariano} also studied unitary or channel discrimination, but all
of them did not consider time-dependent Hamiltonians, which are the
subject of interest here.

A key step towards simplifying Eq.~(\ref{Fx}) is to recognize that
\begin{align}
U_0^\dagger(t_f,t_i) U_1(t_f,t_i) &= \mathcal T
\exp\Bk{-\frac{i}{\hbar}\int_{t_i}^{t_f} dt \Delta H_0(x(t),t)},
\end{align}
where 
\begin{align}
\Delta H_0(x(t),t)\equiv U_0^\dagger(t,t_i)\Delta
H(x(t),t)U_0(t,t_i)
\end{align}
is $\Delta H$ in the \emph{interaction picture} \cite{braginsky}.  In
general, Eq.~(\ref{Fx}) can then be expanded in a Dyson series and
evaluated using perturbation theory \cite{peskin}.  To derive analytic
expressions, however, we shall be more specific about the Hamiltonians
and the initial quantum state.

\section{Force detection with a linear Gaussian system}
Assume that $x$ is a force on a quantum object with position operator
$q$, so that
\begin{align}
\Delta H = -qx,
\end{align}
and the conditional fidelity $F_x$ becomes
\begin{align}
F_x &= \abs{\Avg{\mathcal T
\exp\Bk{\frac{i}{\hbar}\int_{t_i}^{t_f} dt q_0(t)x(t)}}}^2,
\label{Fx2}
\end{align}
with $q_0(t)$ obeying equations of motion under the null hypothesis
$\mathcal H_0$ in the interaction picture. The $\avg{\cdot}$
expression in Eq.~(\ref{Fx2}) is a noncommutative version of the
characteristic functional \cite{gardiner_zoller}. To simplify it,
assume further that $H_0$ consists of terms at most quadratic with
respect to canonical position or momentum operators, such that the
equations of motion are linear and $q_0(t)$ depends linearly on the
initial-time canonical operators. Let $Z(t)$ be a column vector of
canonical position/momentum operators, including $q_0(t)$, that obey
the equation of motion
\begin{align}
\diff{Z(t)}{t} &= G(t)Z(t) + J(t)
\end{align}
under hypothesis $\mathcal H_0$, where $G(t)$ is a drift matrix and
$J(t)$ is a source vector, both consisting of real numbers. $q_0(t)$
can then be written as
\begin{align}
q_0(t) &=  V_q(t,t_i)Z(t_i) + 
\int_{t_i}^t  d\tau V_q(t,\tau)J(\tau),
\end{align}
where $V_q(t,t_i)$ is a row vector and a function of $G(t)$. This gives
\begin{align}
\frac{1}{\hbar}\int_{t_i}^{t_f} dt q_0(t)x(t) =  \kappa^\top Z(t_i) + \phi,
\\
\kappa^\top \equiv \frac{1}{\hbar}\int_{t_i}^{t_f} dt x(t)V_q(t,t_i),
\\
\phi \equiv \frac{1}{\hbar}\int_{t_i}^{t_f} dt x(t)\int_{t_i}^t  d\tau V_q(t,\tau)J(\tau).
\end{align}
With $F_x$ now given by
\begin{align}
F_x &= \abs{\Avg{\mathcal T \exp\Bk{i\kappa^\top Z(t_i) + i\phi}}}^2,
\end{align}
the time-ordering operator becomes redundant:
\begin{align}
F_x &= \abs{\bra{\psi}\exp\Bk{i\kappa^\top Z(t_i)}\ket{\psi}}^2.
\end{align}
This expression can be simplified using the Wigner representation
$W(z,t_i)$ of $\ket{\psi}$, which has the following property
\cite{walls_milburn}:
\begin{align}
\bra{\psi}\exp\Bk{i\kappa^\top Z(t_i)}\ket{\psi}
&= \int dz W(z,t_i) \exp(i\kappa^\top z),
\end{align}
where $z$ is a column vector of phase-space variables.  Assuming
further that $W(z,t_i)$ is Gaussian with mean vector $\bar z$ and
covariance matrix $\Sigma$, we obtain an analytic expression
for $F_x$:
\begin{align}
F_x &= \abs{\int dz W(z,t_i) \exp(i\kappa^\top z)}^2
\\
&= \exp\bk{-\kappa^\top \Sigma \kappa}
\\
&= \exp\Bk{-\frac{1}{\hbar^2}
\int_{t_i}^{t_f} dt \int_{t_i}^{t_f} dt' x(t)\Sigma_q(t,t')x(t')},
\label{Fx_gauss}
\\
\Sigma_q(t,t') &\equiv V_q(t,t_i) \Sigma V_q^\top(t',t_i).
\end{align}
The covariance matrix is given by the Weyl-ordered second moment:
\begin{align}
\Sigma_{jk} &= \frac{1}{2}\Avg{Z_j(t_i)Z_k(t_i) + Z_k(t_i)Z_j(t_i)}
\nonumber\\&\quad
-\Avg{Z_j(t_i)}\Avg{Z_k(t_i)}.
\end{align}
Hence
\begin{align}
\Sigma_q(t,t') &= \frac{1}{2}\Avg{q_0(t)q_0(t')+q_0(t')q_0(t)}
-\Avg{q_0(t)}\Avg{q_0(t')}.
\end{align}
It is interesting to note that the expression given by $-\ln F_x$ in
Eq.~(\ref{Fx_gauss}) coincides with the one proposed in
Refs.~\cite{braginsky,bgkt} as an upper quantum limit on the
force-sensing signal-to-noise ratio, and $4\Sigma_q(t,t')/\hbar^2$ is
equal to the quantum Fisher information in the quantum Cram\'er-Rao
bound for waveform estimation \cite{twc}. The relation of
this expression to the fidelity and the detection error bounds
is a novel result here, however.

If the statistics of $q_0(t)$ can be approximated as
stationary; viz.,
\begin{align}
\Sigma_q(t,t') = \intall \frac{d\omega}{2\pi}
S_q(\omega)\exp[-i\omega(t-t')],
\end{align}
$F_x$ becomes
\begin{align}
F_x &= \exp\Bk{-\frac{1}{\hbar^2}
\intall \frac{d\omega}{2\pi} S_q(\omega)|x(\omega)|^2},
\label{Fx3}
\\
x(\omega) &\equiv \int_{t_i}^{t_f} dt x(t) \exp(i\omega t).
\end{align}
For example, if 
\begin{align}
x(t) = X\cos(\Omega t+\theta)
\end{align}
is a sinusoid,
\begin{align}
F_x &\approx \exp\Bk{-\frac{T}{\hbar^2}S_q(\Omega)X^2},
&
T&\equiv t_f-t_i.
\end{align}
These expressions for the fidelity suggest that, for a given $x(t)$,
there is a fundamental trade-off between force detection performance
and precision in detector position.
%They also suggest an alternative route to the quantum
%Fisher information for waveform estimation studied in Ref.~\cite{twc}
%via a Taylor-series expansion of the fidelity \cite{braunstein}.

\begin{figure}[htbp]
\centerline{\includegraphics[width=0.45\textwidth]{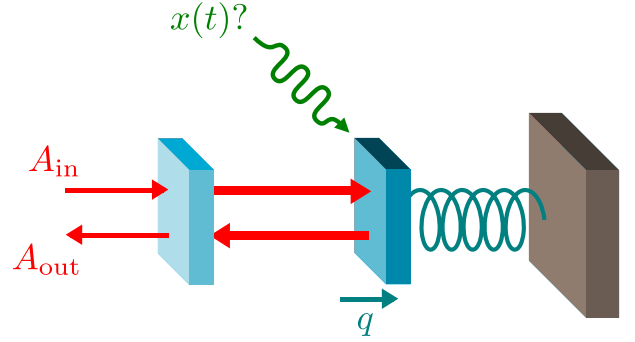}}
\caption{A cavity optomechanical force detector.  An optical cavity
  with a moving mirror is pumped on-resonance with an input field
  $A_\textrm{in}$, and the output field $A_{\rm out}$ is measured to
  infer whether a force $x(t)$ has perturbed the motion of the mirror.}
\label{optomech}
\end{figure}

\section{Optomechanics}
Suppose now that the mechanical object is a moving mirror of an
optical cavity probed by a continuous-wave optical beam, the phase of
which is modulated by the object position and the intensity of which
exerts a measurement backaction via radiation pressure on the object,
as depicted in Fig.~\ref{optomech}. This setup provides a basic and
often sufficient model for more complex optomechanical force
detectors. Let the output field operator under hypothesis $\mathcal
H_j$ be
\begin{align}
  A_{\textrm{out}j}(t) &\approx K_{1}(t) * A_{\textrm{in}}(t)
  +i\mathcal A K_2(t)* q_j(t),
\end{align}
where 
\begin{align}
a(t)*b(t)\equiv \intall d\tau a(t-\tau)b(\tau)
\end{align}
denotes convolution, 
\begin{align}
K_n(t)\equiv \intall \frac{d\omega}{2\pi}K_n(\omega)\exp(-i\omega t)
\end{align}
is an impulse-response function with
\begin{align}
K_{1}(\omega) &\equiv \frac{i\omega+\gamma}{-i\omega+\gamma},
\\
K_{2}(\omega)&\equiv \frac{2\omega_0}{L} \frac{1}{-i\omega+\gamma}
\end{align}
in the
frequency domain, $\mathcal A$ is the input mean field, $\omega_0$ is
the optical carrier frequency, $L$ is the cavity length, and $\gamma$
is the optical cavity decay rate \cite{qnc}. $q_j(t)$ is the position
operator under each hypothesis, which can be written as \cite{qnc}
\begin{align}
q_0(t) &\approx K_3(t)*\hbar K_2(t)*\xi(t),
\\
q_1(t) &\approx K_3(t)*[\hbar K_2(t)*\xi(t) + x(t)],
\end{align}
where $K_3(t)$ is another impulse response function that transfers a
force to the position, 
\begin{align}
\xi\approx \mathcal A^*\Delta
A_\textrm{in}(t) +\mathcal A\Delta A_\textrm{in}^\dagger(t)
\end{align}
is the backaction noise, and the transient solutions are assumed to
have decayed to zero. Defining 
\begin{align}
K_4(t)\equiv K_3(t)*K_2(t),
\end{align}
such that the position power spectral density is 
\begin{align}
S_q(\omega) =
\hbar^2|K_4(\omega)|^2 S_\xi(\omega),
\end{align}
we obtain
\begin{align}
F_x &= \exp\Bk{-\intall \frac{d\omega}{2\pi} 
S_\xi(\omega)|K_4(\omega)x(\omega)|^2}.
\label{Fx4}
\end{align}

The backaction noise $\xi$ that appears in the output field, in
addition to the shot noise in $A_{\textrm{in}}$, can limit the
detection performance at the so-called standard quantum limit
\cite{braginsky,bvt,caves,caves1985}.  This does not seem to agree
with the fundamental quantum limits in terms of Eq.~(\ref{Fx4}), which
suggest that increased fluctuations in $q_0(t)$ due to $\xi(t)$ can
improve the detection.  Fortunately, it is now known that the
backaction noise can be removed from the output field
\cite{yuen1983,ozawa1988,qnc,qmfs,caniard,julsgaard,hammerer,wasilewski,klmtv}.
One method, called quantum-noise cancellation (QNC), involves passing
the optical beam through another quantum system that has the effective
dynamics of an optomechanical system with negative mass
\cite{qnc,qmfs,julsgaard,hammerer,wasilewski}. With the backaction
noise removed, the output fields become
\begin{align}
A_{\textrm{out}0}(t) &\approx K_1(t)*A_\textrm{in}(t),
\label{Aout0}\\
A_{\textrm{out}1}(t) &\approx K_1(t)*A_\textrm{in}(t) + i\mathcal A 
K_2(t)*K_3(t)*x(t).
\label{Aout1}
\end{align}
If the phase quadrature of $A_{\textrm{out}j}(t)$ is measured by
homodyne detection, the outputs can be written as
\begin{align}
y_0(t) &\approx \eta(t),
\label{y0}
\\
y_1(t) &\approx \eta(t) + K_2(t)*K_3(t)*x(t),
\label{y1}\\
\eta(t) &\equiv 
\frac{1}{2i|\mathcal A|^2}[\mathcal A^* 
K_1(t)*\Delta A_\textrm{in}(t)-\mathcal A K_1^*(t)*\Delta A_\textrm{in}^\dagger(t)].
\end{align}
The power spectral densities of $\xi(t)$ and $\eta(t)$ satisfy an
uncertainty relation \cite{braginsky}:
\begin{align}
S_\xi(\omega)S_\eta(\omega) &\ge \frac{1}{4}.
\label{uncertain}
\end{align}
The detection problem described by Eqs.~(\ref{y0}) and (\ref{y1})
becomes a classical one with additive Gaussian noise, a scenario that
has been studied extensively in gravitational-wave detection
\cite{flanagan,flanagan2}.

\section{Error bounds for deterministic waveform detection}
Suppose that $x(t)$ is known \textit{a priori}. It is then well known
that the error probabilities for the detection problem described by
Eqs.~(\ref{y0}) and (\ref{y1}) using a likelihood-ratio test are
\cite{vantrees}
\begin{align}
P_{10,\textrm{hom}} &= \frac{1}{2}\erfc\bk{\sigma + \frac{\lambda}{4\sigma}},
\\
P_{01,\textrm{hom}} &= \frac{1}{2}\erfc\bk{\sigma - \frac{\lambda}{4\sigma}},
\end{align}
where 
\begin{align}
\erfc u &\equiv \frac{2}{\sqrt{\pi}}\int_{u}^\infty
dv\exp(-v^2),
\end{align}
$\lambda$ is the threshold in the likelihood-ratio
test, which can be adjusted according to the desired criterion, and
$\sigma$ is a signal-to-noise ratio given by 
\begin{align}
\sigma^2 \approx
\frac{1}{8}\intall \frac{d\omega}{2\pi} 
\frac{|K_4(\omega)x(\omega)|^2}{S_\eta(\omega)}
\end{align}
for a long observation time relative to the duration of $x(t)$ plus
the decay time of $K_4(t)$.  To compare homodyne detection with the
quantum limits, suppose that the duration of $x(t)$ is long and
$\sigma^2$ increases at least linearly with $T$, so that we can define
an error exponent as the asymptotic decay rate of an error probability
in the long-time limit. For simplicity, we consider here only the
exponent of the higher error probability:
\begin{align}
\Gamma &\equiv -\lim_{T\to\infty}\frac{1}{T} \ln \max\BK{P_{10},P_{01}}.
\end{align}
Although this asymptotic limit may not be relevant to
gravitational-wave detectors in the near future, the error
probabilities for which are anticipated to remain high, we focus on
this limit to obtain simple analytic results, which allow us to gain
useful insight into the fundamental physics. More precise calculations
of error probabilities are more tedious but should be straightforward
following the theory outlined here.

For homodyne detection, the error exponent is
\begin{align}
\Gamma_\textrm{hom} &= \frac{\sigma^2}{T}
=\frac{1}{8T}\intall \frac{d\omega}{2\pi}
\frac{|K_4(\omega)x(\omega)|^2}{S_\eta(\omega)}.
\end{align}
The quantum limit, on the other hand, is
\begin{align}
&\quad -\lim_{T\to\infty} \frac{1}{T}\ln \max\BK{P_{10},P_{01}}
\nonumber\\
&\le 
-\lim_{T\to\infty} \frac{1}{T}\ln \max_{P_0} P_{e} 
\le 
-\lim_{T\to\infty} \frac{1}{T}\ln F\equiv \Gamma_F,
\end{align}
which gives
\begin{align}
\Gamma_F &= \frac{1}{T}\intall \frac{d\omega}{2\pi}
S_\xi(\omega) |K_4(\omega)x(\omega)|^2.
\end{align}
Using the uncertainty relation between $S_\xi$ and $S_\eta$ in
Eq.~(\ref{uncertain}), it can be seen that
\begin{align}
\Gamma_{\rm hom} \le \frac{\Gamma_F}{2},
\end{align}
that is, the homodyne error exponent is at most half the optimal
value.  This fact is well known in the context of coherent-state
discrimination \cite{helstrom,cook,wittmann,tsujino}. The
suboptimality of homodyne detection here should be contrasted with the
conclusion of Ref.~\cite{twc}, which states that homodyne detection
together with QNC are sufficient to achieve the quantum limit for the
task of waveform estimation.

\begin{figure}[htbp]
\centerline{\includegraphics[width=0.45\textwidth]{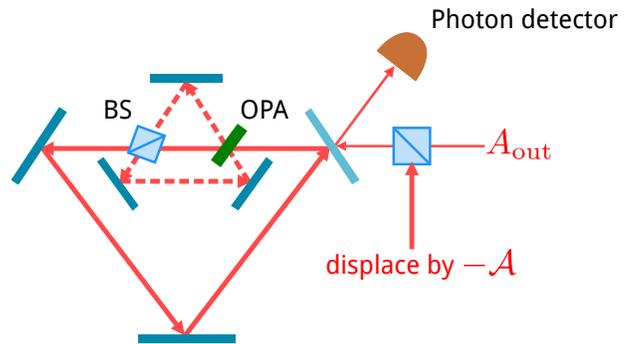}}
\caption{An integrated QNC-Kennedy receiver.  The output field
  $A_\textrm{out}$ from the optomechanical force detector in
  Fig.~\ref{optomech} is displaced by $-\mathcal A$ and then passed
  through an optical setup that removes the measurement backaction
  noise. The dash arrows represent a red-detuned optical cavity mode
  that mimics a negative-mass oscillator and interacts with the
  optical probe field via a beam splitter (BS) and a two-mode optical
  parametric amplifier (OPA). Details of how this setup works can be
  found in Refs.~\cite{qnc,qmfs}. If the field $A_{\rm in}$ is in a
  coherent state, the final output field should be in a vacuum state
  under the null hypothesis $\mathcal H_0$. Any photon detected at the
  output indicates that $\mathcal H_1$ must be true.}
\label{qnc_kennedy}
\end{figure}

To see how one can get closer to the quantum limits, let's go back to
Eqs.~(\ref{Aout0}) and (\ref{Aout1}). Observe that, if the input field
is in a coherent state, the output field is also in a coherent state
(in the Schr\"odinger picture) under each hypothesis. This means that
existing results for coherent-state discrimination can be used to
construct an optimal receiver.  The Kennedy receiver, for example,
displaces the output field so that it becomes vacuum under $\mathcal
H_0$ and then detects the presence of any output photon
\cite{helstrom}. Any detected photon means that $\mathcal H_1$ must be
true. Deciding on $\mathcal H_0$ if no photon is detected and
$\mathcal H_1$ otherwise, the false-alarm probability $P_{10}$ is
zero, while the miss probability is the probability of detecting no
photon given $\mathcal H_1$, or
\begin{align}
P_{01,\textrm{Ken}} 
&= \exp \Bk{-
\int_{t_i}^{t_f}dt \abs{\mathcal AK_2(t)*K_3(t)*x(t)}^2}.
\end{align}
For a long observation time with $S_\xi = |\mathcal A|^2$ for a
coherent state,
\begin{align}
P_{01,\textrm{Ken}}&\approx
\exp \Bk{-\intall \frac{d\omega}{2\pi} S_\xi|K_4(\omega)x(\omega)|^2}
= F,
\end{align}
which makes the Kennedy receiver optimal under the Neyman-Pearson
criterion in the case of $P_{10} = 0$ according to Eq.~(\ref{P01}) and
also achieve the optimal error exponent:
\begin{align}
\Gamma_\textrm{Ken} &= -\lim_{T\to\infty}\frac{1}{T}\ln P_{01,\textrm{Ken}}
= \Gamma_F.
\end{align}
The Kennedy receiver can be integrated with the QNC setup; an example
is shown in Fig.~\ref{qnc_kennedy}.  The Dolinar receiver, which
updates the displacement field continuously according to the
measurement record, can further improve the average error probability
slightly to saturate the lower limit given by Eq.~(\ref{Pe})
\cite{helstrom,cook}. Other more recently proposed receivers may also
be used here to beat the homodyne limit \cite{wittmann,tsujino}.

\section{Error bounds for stochastic waveform detection}
Consider now a stochastic $x(t)$, which should be relevant to the
detection of stochastic backgrounds of gravitational waves
\cite{regimbau}.  Since $F_x$ is Gaussian,
\begin{align}
F &= \int Dx P[x] \exp\Bk{-\frac{1}{\hbar^2}\int dt dt' x(t)\Sigma_q(t,t')x(t')}
\end{align}
can be computed analytically if the prior $P[x]$ is also
Gaussian. Here we shall use a discrete-time approach and take the
continuous limit at the end of our calculations. If $x(t)$ is a
zero-mean Gaussian process with covariance
\begin{align}
\Sigma_x(t,t') \equiv
\mathbb E[x(t)x(t')],
\end{align}
it can be discretized as 
\begin{align}
x &\equiv
(x_0,\dots,x_{N-1})^\top,
\\
Dx P[x] &\approx dx_0\dots dx_{N-1}
\frac{1}{\sqrt{(2\pi)^N\det \Sigma_x}}
\nonumber\\&\quad\times
\exp\bk{-\frac{1}{2}x^\top \Sigma_x^{-1} x},
\\
\Sigma_x&\equiv \mathbb E(x x^\top).
\end{align}
The fidelity then becomes a finite-dimensional Gaussian integral:
\begin{align}
F &\approx \int dx_0\dots dx_{N-1}
\frac{1}{\sqrt{(2\pi)^N\det \Sigma_x}}
\nonumber\\&\quad\times
\exp\bk{-\frac{1}{2}x^\top \Sigma_x^{-1} x-\frac{\delta t^2}{\hbar^2} x^\top \Sigma_q x}
\\
&= \sqrt{\frac{\det (\Sigma_x^{-1}+2\delta t^2\Sigma_q/\hbar^2)^{-1}}{\det \Sigma_x} }
\\
&= \Bk{\det\bk{I+\frac{2\delta t^2}{\hbar^2}\Sigma_q\Sigma_x}}^{-1/2}
\\
&= \exp\Bk{-\frac{1}{2}\trace \ln\bk{I+\frac{2\delta t^2}{\hbar^2}\Sigma_q\Sigma_x}}
\\
&= \exp\bk{-\frac{1}{2}\sum_\omega \ln\lambda_\omega},
\end{align}
where $\lambda_\omega$ are the eigenvalues of the matrix 
\begin{align}
C\equiv
I+\frac{2\delta t^2}{\hbar^2}\Sigma_q\Sigma_x.
\end{align}
If $\Sigma_q(t,t')$ and
$\Sigma_x(t,t')$ are both stationary; viz.,
\begin{align}
\Sigma_q(t,t') &=
\sigma_q(t-t'),
\\
\Sigma_x(t,t') &= \sigma_x(t-t'),
\end{align}
they can be modeled as circulant matrices in discrete time, so that
$C$ is also circulant, with eigenvalues given by the discrete Fourier
transform of a row or column vector of the matrix. Taking the
continuous-time limit using
\begin{align}
\sum_\omega \to T\intall \frac{d\omega}{2\pi},
\end{align}
we get
\begin{align}
F &= \exp(-\Gamma_F T),
\\
\Gamma_F &= \frac{1}{2}\intall \frac{d\omega}{2\pi}
\ln\Bk{1+\frac{2}{\hbar^2}S_q(\omega)S_x(\omega)},
\\
S_q(\omega) &\equiv \intall dt \sigma_q(t)\exp(i\omega t),
\\
S_x(\omega) &\equiv \intall dt \sigma_x(t)\exp(i\omega t).
\end{align}
This fidelity expression can then be used in the detection error
bounds.

For homodyne detection, the error exponent is more complicated for
stochastic waveform detection and given by the Chernoff distance
\cite{vantrees3,[{}] [{, Chap.~10.}] levy}:
\begin{align}
&\quad \Gamma_{\rm hom} 
\nonumber\\
&=
\sup_{0\le s\le 1}
\frac{1}{2}\intall \frac{d\omega}{2\pi}
\ln \frac{1+(1-s)|K_4(\omega)|^2S_x(\omega)/S_\eta(\omega)}
{[1+|K_4(\omega)|^2S_x(\omega)/S_\eta(\omega)]^{1-s}}.
\end{align}
The performance of homodyne detection
relative to the quantum limits then depends on the specific form of
$|K_4(\omega)|^2S_x(\omega)$. The Kennedy receiver, on the other hand,
is still applicable here, as the output is still a coherent state
under $\mathcal H_0$. The false-alarm probability is still zero, and
the miss probability is now
\begin{align}
P_{01,\textrm{Ken}} &\approx
\mathbb E\exp \Bk{-
\intall \frac{d\omega}{2\pi} S_\xi(\omega)
|K_4(\omega)x(\omega)|^2}\approx F,
\end{align}
which means that the Kennedy receiver remains optimal, both in terms
of the Neyman-Pearson criterion in the case of $P_{10} = 0$ and the
error exponent. Whether other receivers can do even better and
saturate the other quantum bounds is a more difficult question, as the
output field under $\mathcal H_1$ is now in a mixed state and the
fidelity lower bounds may not be achievable. 

The use of Kennedy or Dolinar receivers assumes coherent states at the
output, which is the case only if the backaction noise cancellation is
complete and quantum shot noise in the input beam is the only source
of noise at the output.  Although such assumptions are highly
idealistic, especially for current gravitational-wave detectors, the
ideal scenario shows that the quantum bounds proposed here are in
principle achievable using known optics technology.  Optimal
discrimination of squeezed or other Gaussian states remains a topic of
current research \cite{tan,guha,pirandola} and may be useful for
future gravitational-wave detectors that use squeezed light
\cite{ligo2011}. Generalization of the results here to multi-waveform
discrimination should also be useful for gravitational-wave astronomy
\cite{flanagan,flanagan2} and may be done by following
Refs.~\cite{helstrom,ykl,bondurant}.

\section{Outlook}
Now that quantum limits to waveform detection have been discovered,
the natural next question to ask is how they can be approached in
practice. In the case of optomechanical force detection, the
requirements are quantum shot noise as the only source of noise at the
output and an appropriate receiver, such as the Kennedy receiver. A
proof-of-concept experimental demonstration of waveform detection
approaching the shot-noise limits based on Eq.~(\ref{Fx4}) should be
well within reach of current quantum optics technology. To demonstrate
the trade-off between force detection performance and detector
localization suggested by Eq.~(\ref{Fx3}) with optomechanics, however,
can be much more challenging, as it would require quantum backaction
noise to dominate the detector position fluctuation but become
negligible in the output via QNC. A more promising candidate for this
demonstration is atomic spin ensembles, with which
backaction-noise-canceled magnetometry has already been realized
\cite{wasilewski}. The likelihood-ratio formulas derived in
Ref.~\cite{hypothesis} should be used in practice instead of the
ideal-case decision rules discussed here to account for any excess
noise.

%Regardless of the immediate prospect of a demonstration or the
%relevance of the proposed quantum limit to current applications, there
%is no question that the progress of science and technology depends on
%increasingly precise measurements, and quantum mechanics will
%eventually become a major concern. When that happens, the quantum
%limit proposed here is envisioned to play a fundamental role in the
%modeling and design of all quantum detectors.

In terms of potential further theoretical work, it should be useful to
generalize beyond the assumptions of scalar waveform, linear Gaussian
systems, optical coherent states, stationary processes, and long
observation time used here. The fidelity expressions derived here may
also be useful for the study of waveform estimation \cite{twc}, either
as an alternative way of deriving the quantum Fisher information via a
Taylor-series expansion \cite{braunstein} or used directly in the
quantum Ziv-Zakai bound \cite{qzzb}.

From a more conceptual point of view, this study, together with the
earlier work on waveform estimation \cite{twc}, shows that the
concepts of states, effects, and operations fade into background when
dealing with dynamical quantum information systems, and
\emph{multi-time} quantum statistics, through the use of Heisenberg or
interaction picture, take the center stage. It may be interesting to
explore whether this perspective has any relevance to other dynamical
quantum information systems, such as quantum computers \cite{nielsen},
and the study of quantum correlations \cite{horodecki}.

\section*{Acknowledgments}
Discussions with Brent Yen, Andy Chia, Carlton Caves, and Howard
Wiseman are gratefully acknowledged.  This material is based on work
supported by the Singapore National Research Foundation under NRF
Grant No.~NRF-NRFF2011-07.

%\bibliography{research}


\end{document}